\documentstyle[aps,preprint,epsf]{revtex}
\newcommand{\beq}{\begin{equation}}
\newcommand{\eeq}{\end{equation}}
\begin{document}
\draft
\tightenlines

\title{ Fermion Condensation Quantum Phase Transition versus
Conventional Quantum Phase Transitions}

\author{ V.R. Shaginyan$^{a,b}$ \footnote{E--mail:
vrshag@thd.pnpi.spb.ru}, J.G. Han$^b$, J. Lee$^b$}
\address{$^a$Petersburg Nuclear Physics Institute,
Gatchina, 188300, Russia\\
$^{b}$CAPST, SungKyunKwan University 300, Chun-Chun-Dong Jangan-gu
Suwon 44-746, Korea} \maketitle

\begin{abstract}

The main features of fermion condensation quantum phase transition
(FCQPT), which are distinctive in several aspects from that of
conventional quantum phase transition (CQPT), are considered.  We
show that in contrast to CQPT, whose physics in quantum critical
region is dominated by thermal and quantum fluctuations and
characterized by the absence of quasiparticles, the physics of a
Fermi system near FCQPT or undergone FCQPT is controlled by the
system of quasiparticles resembling the Landau quasiparticles.
Contrary to the Landau quasiparticles, the effective mass of these
quasiparticles strongly depends on the temperature, magnetic fields,
density, etc. This system of quasiparticles having  general
properties determines the universal behavior of the Fermi system in
question.  As a result, the universal behavior persists up to
relatively high temperatures comparatively to the case when  such a
behavior is determined by CQPT. We analyze striking recent measurements
of specific heat, charge and heat transport used to study the nature of
magnetic field-induced QCP in heavy-fermion metal CeCoIn$_5$ and show
that the observed facts are in good agreement with our scenario based
on FCQPT and certainly seem to rule out the critical fluctuations
related with CQPT. Our general consideration suggests that FCQPT and the
emergence of novel quasiparticles near and behind FCQPT and resembling
the Landau quasiparticles are distinctive features intrinsic to
strongly correlated substances.

\end{abstract}


It is generally accepted that the fundamental physics that gives rise
to the high-$T_c$ superconductivity and non-Fermi liquid (NFL) behavior
with a recovery of the Landau-Fermi liquid (LFL) behavior under
the application of magnetic fields observed in heavy-fermion (HF)
metals and high-$T_c$ compounds is controlled by quantum phase
transitions. This has made quantum phase transitions a subject of
intense current interest, see e.g. \cite{sac,voj}.

A quantum phase transition is
driven by control parameters such as composition, density or magnetic
fields and takes place at a quantum critical point (QCP) when
temperature $T=0$. QCP separates an ordered
phase generated by quantum phase transition from a disordered phase.
It is expected that the universal behavior is only observable if the
system in question is very near to QCP, for example, when the
correlation length is much larger than microscopic length scales.
Quantum phase transitions of this sort are quite common, and we shall
label such  quantum phase transitions as conventional quantum phase
transitions (CQPT). In the case of CQPT, the physics is dominated by
thermal and quantum fluctuations of the critical state, which is
characterized by the absence of quasiparticles.
It is believed that the absence of quasiparticle-like
excitations is the main cause of the NFL behavior
and other types of the critical behavior in the quantum
critical region. Basing on the assumption of scaling
related to the divergency of the correlation length, one can
construct the critical contribution to the free energy and evaluate
the corresponding properties such as critical exponents, the NFL
behavior, etc. \cite{sac,voj}. Moving along this way, one may expect
difficulties. For example, having the only critical contribution, one
has to describe at least two types of the behavior exhibited by
different HF metals, see e.g.  \cite{geg1,alp,shag4}. Note that HF
metals are three-dimensional structures,
see e.g. \cite{geg1,kad,kadw,geg} and, thus, the
type of behavior cannot be related to the dimension. The critical
behavior observed in measurements on HF metals takes place up to
rather high temperatures comparable with the effective Fermi
temperature $T_k$. For example, the thermal expansion coefficient
$\alpha(T)$ measured on CeNi$_2$Ge$_2$ shows a $1/\sqrt{T}$ divergence
over more than two decades in temperature from 6 K down to at least
50 mK \cite{geg1}.  It is hardly possible to understand such a
behavior basing on the assumption of scaling when the correlation
length is to be much larger than microscopic length scales.
Obviously, such a situation can only take place at $T\to0$. At $T\sim
T_k$, this macroscopically large correlation length must be destroyed
by thermal fluctuations.  The next problem is related to explanations
of recovery of the LFL behavior
under applied magnetic fields $B$
observed in HF metals. At $T\to0$, the magnetic field dependence of
the coefficient $A(B)$, the Sommerfeld coefficient $\gamma(B)$ and
$\chi(B)$ in the resistivity $\rho=\rho_0+\Delta\rho$,
$\Delta\rho=A(B)T^2$, specific heat, $C/T=\gamma(B)$, and magnetic
susceptibility, $\chi(B)$, shows that $A(B)\sim\gamma(B)^2$ and
$A(B)\sim\chi(B)^2$, so that the Kadowaki-Woods ratio,
$K=A(B)/\gamma^2(B)$ \cite{kadw}, is conserved \cite{geg}. Such a
universal behavior is hardly possible to explain within the picture
assuming the absence of quasiparticles which takes place near QCP of
the corresponding CQPT. As a consequence, for example, these facts are
in variance to the spin-density-wave scenario \cite{geg}
and the renormalization group treatment of quantum criticality
\cite{mill}. Moreover, striking recent measurements
of specific heat, charge and heat transport used to study the nature of
magnetic field-induced QCP in heavy-fermion metal CeCoIn$_5$
 \cite{bi,pag1} certainly seem to disagree with descriptions based on
CQPT.

In this Letter, we consider the main features of fermion
condensation quantum phase transition (FCQPT), which is distinctive
in several aspects from CQPT. We show that in contrast to
CQPT, whose physics in the quantum critical region is dominated by
thermal or quantum fluctuations and characterized by the absence of
quasiparticles, the physics of a Fermi system near FCQPT or undergone
FCQPT is controlled by the system of quasiparticles resembling the
Landau quasiparticles. Since FCQPT is characterized by the
superconducting order parameter $\kappa({\bf p})$ \cite{ms},
the critical region of fluctuations of $\kappa({\bf p})$ is
extremely narrow at $T\to0$ and therefore it gives a negligible small
critical contribution to the free energy.
Contrary to the Landau quasiparticles, the effective mass of these
quasiparticles strongly depends on the temperature, magnetic
fields, density,  etc. This system of quasiparticles having
general properties determines the universal behavior of the Fermi
system in question at finite temperatures and in magnetic fields
being the main cause of the NFL behavior and other types of the
critical behavior. As a result, the universal behavior persists up to
relatively high temperatures and magnetic fields comparatively to the
case when such a behavior were determined by CQPT. We analyze
striking recent measurements of specific heat, charge and heat
transport which where used to study the nature of magnetic
field-induced QCP in heavy-fermion metal
CeCoIn$_5$ \cite{bi,pag1} and show that the
observed facts are in good agreement with our scenario based on
FCQPT.

In the LFL theory, the existence of low-energy elementary excitations
is the general property of a Fermi liquid and these
excitations are represented by quasiparticles \cite{lan}.
The quasiparticle distribution function $n({\bf p},T)$ is
given by the equation \begin{equation} \frac{\delta \Omega}{\delta
n({\bf p},T)} =\varepsilon ({\bf p},T)-\mu (T)-T\ln \frac{1-n({\bf
p},T)}{n({\bf p},T)}=0.  \end{equation} The function $n({\bf p},T)$
depends on the momentum ${\bf p}$ and the temperature $T$.  Here
$\Omega=E-TS-\mu N$  is the thermodynamic potential, and $\mu $ is
the chemical potential, while $\varepsilon ({\bf p},T)$,
\begin{equation}
\varepsilon ({\bf p},T)=\frac{\delta E[n(p)]}{\delta n({\bf p},T)},
\end{equation}
is the quasiparticle energy. This energy is a functional of $n({\bf
p},T)$ just like the total energy $E[n(p)]$, entropy $S[n(p)]$ and
the other thermodynamic functions. The entropy $S[n(p)]$ is given by
the familiar expression \begin{equation} S[n(p)]=-2\int \left[ n({\bf
p},T)\ln n({\bf p},T)+(1-n({\bf p},T)) \ln (1-n({\bf p},T))\right]
\frac{d{\bf p}}{(2\pi )^3}, \end{equation} which results from purely
combinatorial considerations. Eq. (1) is usually presented as the
Fermi-Dirac distribution \begin{equation} n({\bf p},T)=\left\{ 1+\exp
\left[ \frac{(\varepsilon ({\bf p},T)-\mu )}T \right] \right\} ^{-1}.
\end{equation} At $T\to 0$, one gets from Eqs. (1), (4) the standard
solution $n_F({\bf p},T\to 0)\to \theta (p_F-p)$, with $\varepsilon
(p\simeq p_F)-\mu =p_F(p-p_F)/M^{*}_L$, where $p_F$ is the Fermi
momentum, $\theta (p_F-p)$ is the step function, and $M^{*}_L$ is the
Landau effective mass \cite{lan} \beq
\frac{1}{M^*_L}=\frac{1}{p}
\frac{d\varepsilon(p,T\to0)}{dp}|_{p=p_F}.
\eeq
It is implied that in the case of LFL $M^{*}_L$ is
positive and finite at the Fermi momentum $p_F$. As a result, the
$T$-dependent corrections to $M^{*}_L$, to the quasiparticle energy
$\varepsilon (p)$, and to other quantities, start with $T^2$-terms
being approximately temperature independent.

There exist special solutions of Eq. (1)
associated with the so-called fermion condensation \cite{ks,ksk}.
Being continuous and satisfying the inequality $0<n_0({\bf p})<1$
within some region in $p$, such solutions $n_0({\bf p})$ admit a
finite limit for the logarithm in Eq.  (1) at $T\rightarrow 0$
yielding  \cite{ks,ksk} \begin{equation} \varepsilon({\bf
p})-\mu=0,\quad \mbox{if} \quad 0<n_0({\bf p})<1;\,p_i\leq p\leq p_f,
\end{equation} where  $\varepsilon({\bf p})$ is given by Eq. (2).  At
$T=0$, Eq. (6) defines a new state of electron liquid with FC
\cite{ks,ksk,vol}, which is characterized by a flat spectrum in the
$(p_f-p_i)$ region, and which can strongly
influence measurable quantities up to temperatures
$T\ll T_f$. Here $T_f$ stands for the temperature
where the fermion condensate effects disappear,
$T_f/\varepsilon_F\sim (p_f-p_i)/p_F$, \cite {ksk}.
Here $\varepsilon_F\sim p_F^2/M^*_L$ is the Fermi energy.
Note that at $T\ll T_f$ the occupation numbers of quasiparticles
is approximately temperature independent,
$n({\bf p},T)\simeq n_0({\bf p})$, with $n_0({\bf p})$	being given
by Eq. (6).
At $T=0$ in the state with FC, the order parameter
coincides with the order parameter of
superconducting state
$\kappa ({\bf p})=\sqrt{(1-n_0({\bf p}))n_0({\bf p})}$ and
has finite values in the $(p_f-p_i)$ region.
While, the maximum value of the superconducting gap $\Delta_1\to 0$
in this region and the transition temperature $T_c\to0$, provided
that the pairing interaction tends to zero.  Such a state can be
considered as superconducting, with an infinitely small value of
$\Delta_1$, so that the entropy $S(T=0)$ of this state is equal to
zero \cite{ms,ksk,ks1}.

When $p_f\to p_i\to p_F$ the flat part vanishes, and Eq. (6)
determines QCP at which the effective mass $M^*$ diverges
due to density and spin fluctuations
\cite{ksk,ksem,ksz,zks,khod,khod1,shag1}.
FCQPT manifests itself in the divergence of the quasiparticle
effective mass $M^*$ as the density $x$ tends to the critical
density $x_{FC}$, or the distance $r=(x-x_{FC})\to 0$ being positive
\cite{khod,khod1,shag1} \begin{equation} M^*\propto
\frac{1}{x-x_{FC}}\propto \frac{1}{r}.	\end{equation} As long as the
effective mass $M^*$ is finite, the system exhibits the LFL behavior
at low temperatures $T\sim T^*(x)\propto|x-x_{FC}|^2$ \cite{shag1}.
The behavior of $M^*$ given by Eq. (7) is in good agreement with
experimental facts obtained both in measurements on 2D Fermi systems
such as electron gas and $^3$He \cite{cas,skdk,cas1} and
recent calculations \cite{krot,ying,tant}.
We note that at $x\to x_{FC}$, the quasiparticle renormalization
factor $z$ remains approximately constant and
the divergence of the effective mass $M^*$ is
not related to vanishing $z$, see e.g. \cite{khod,ying,khorep}.
Thus, Eq. (6) possesses non-trivial solutions at
the critical quantum point $x=x_{FC}$ as soon as the kinetic
energy $E_k\sim p_F^2/M^*$ becomes
frustrated and the effective inter-electron interaction, or the
Landau amplitude, being sufficiently large,
start to determine the occupation numbers $n({\bf
p})$ which deliver the minimum value to the energy $E[n(p)]$.  As a
result, the occupation numbers $n({\bf p})$ become variational
parameters and Eq. (6) has non-trivial solutions $n_0({\bf p})$,
because the energy $E[n(p)]$ can be lowered by alteration of the
occupation numbers. Thus, within the region $p_i<p<p_f$, the solution
$n_0({\bf p})$ deviates from the Fermi step function $n_F({\bf p})$
in such a way that the energy $\varepsilon({\bf p})$ stays constant,
while outside this region $n_0({\bf p})$ coincides with
$n_F({\bf p})$ \cite{ks,ksk}. In response to this, the system becomes
divided into two quasiparticle subsystems: the first subsystem in the
$(p_f-p_i)$ range is characterized  by quasiparticles with the
effective mass $M^*_{FC}\to \infty$, while the second one is occupied
by quasiparticles with finite mass $M^*_L$ and momenta $p<p_i$.
It is seen from Eq. (7) that in the absence of FCQPT the effective
mass could become negative at $r<0$.
Thus, forming FC below the critical point $x_{FC}$, the system
escapes the possibility to be in meaningless states with negative
values of the effective mass.  We note, that a formation of the flat
part of the spectrum has been observed in \cite{dzy,irk} and obtained
within exactly solvable models as well \cite{lid}.

At $\Delta_1\to 0$, the critical temperature $T_c\to 0$.
We see that the ordered phase with the order parameter
$\kappa({\bf p})$ and FC can exist only at $T=0$,
and the state of electron liquid with FC disappears at $T>0$
\cite{ms,ks1}. Therefore, FCQPT is not the endpoint of a line of
finite-temperature phase transitions. This conclusion is in
accordance with Eq. (1) which does not admit the existence of the
flat part of spectrum at finite temperatures. FCQPT is
driven by the divergency of the effective mass of quasiparticles.
To put it differently, at QCP when $x\to x_{FC}$, the
effective mass $M^*$ diverges due to both density and spin
fluctuations, and in its turn, making the kinetic energy be
frustrated, drives FCQPT with the superconducting order parameter
$\kappa({\bf p})$.  Importantly, that the density and spin fluctuations
in the question are not the critical fluctuations and the divergency
of the effective mass is not a result
of the critical fluctuations or vanishing the quasiparticle
renormalization factor $z$ \cite{ksz,khod,shag1,ying,khorep}.
We remark that in the case of the
superconducting phase transition when $T_c/\varepsilon_F\ll 1$, the
critical region of the critical fluctuations of the order parameter
is so narrow that is almost irrelevant from the experimental point of
view \cite{lanl1}. Thus, we can conclude that at finite temperatures
$T>T_c$ the physics of Fermi system near FCQPT or behind the critical
point of FCQPT is controlled by the system of quasiparticles
resembling the Landau quasiparticles.  Therefore, at $T_c\to0$, the
area of applications of the quasiparticle scenario extends
practically to $T=0$ because the critical region becomes very narrow
and eventually vanishes.

Assume that $T_c\to 0$ and $r<0$, then
at finite temperatures $T\ll T_f$, the occupation numbers in the
region $(p_f-p_i)$ are still determined by Eq. (6), and the
quasiparticle system
becomes divided into two quasiparticle subsystems:  the first
subsystem is occupied by normal quasiparticles with the finite
effective mass $M_L^{*}$ independent of $T$ at momenta $p<p_i$, while
the second subsystem in the $(p_f-p_i)$ range is characterized by the
quasiparticles with the effective mass $M_{FC}^{*}(T)$
\cite{ms,ksk,ks1} \begin{equation} M_{FC}^{*}\simeq
p_F\frac{p_f-p_i}{4T}.	\end{equation} There is an energy scale $E_0$
separating the slow dispersing low energy part, related to the
effective mass $M_{FC}^{*}$, from the faster dispersing relatively
high energy part, defined by the effective mass $M_L^{*}$. It follows
from Eq. (8) that $E_0$ is of the form \cite{ms}
\begin{equation}E_0\simeq 4T.\end{equation}

At $r<0$, the Fermi system  can be viewed as a
strongly correlated one demonstrating the NFL behavior even at low
temperatures as it follows from Eqs. (8) and (9). At $T\to0$, it
behaves as a Fermi system moving towards a quantum critical point, for
example, the effective mass  diverges,
$M^*_{FC}\propto 1/T$, while the Gr\"uneisen
ratio, ${\rm\Gamma(T)}=\alpha(T)/C(T)$, diverges algebraically,
${\rm\Gamma(T)}\propto 1/\sqrt{T}$ \cite{alp,shag4}.  Here, $\alpha(T)$
is the thermal expansion coefficient and $C(T)$ is the specific heat.
In fact, at $T\to0$, the system approaches the FC quantum critical line
from above, and its behavior is controlled by the system of
quasiparticles with the effective masses $M^*_{FC}$ and $M_L$. At
$T\to0$, the Fermi system in question is undergone a weakly
first-order quantum phase transition because the entropy is not a
continuous function at $T=0$:  $\delta S=S(T>0)-S(T=0)$ is finite
since $S(T=0)=0$, and at $0<T\ll T_f$, the entropy possesses the
finite contribution coming from the occupations numbers $n_0({\bf
p})$, see Eqs. (3) and (6).  On the other hand, according to the well
known inequality, $\delta Q\leq T\delta S$, the heat $\delta Q$ of
the transition is equal to zero, because $T=T_c\to 0$.	In the same
way, other thermodynamic functions such as the thermodynamic
potential $\Omega$ and the free energy, $F=E-TS$, are continuous
functions at $T\to0$.  Actually, FC cannot survive at $T\to0$ because
of the degeneracy of the FC spectrum, being absorbed by phase
transitions removing the degeneracy. For example, FCQPT can be
absorbed by the superconducting phase transition, see below. In that
case, the critical temperature $T_c$ becomes finite, and at $T\to
T_c$, we have the continues phase transition. It is pertinent to note
that contrary to the considered case, a conventional quantum phase
transition, at its quantum critical line, makes Gr\"uneisen ratio
${\rm\Gamma(T)}$ diverge at most logarithmically,
${\rm\Gamma(T)}\propto \pm\log T$, \cite{zhu}.

At $T\geq 0$ and $r=(x-x_{FC})>0$, the system is on the disordered side
and the effective mass given by Eq. (7) becomes finite. As a result,
the kinetic energy comes into a play  and
makes the flat part vanish. Obviously, at $T=0$, Eq. (6) has only the
trivial solution $\varepsilon (p=p_F)=\mu $, and the quasiparticle
occupation numbers are given by the step function, $n_F({\bf p})=\theta
(p_F-p)$.
At $\varepsilon_F\gg T>T^*(x)$ and $|x-x_{FC}|/x_{FC}\ll 1$,
the effective mass $M^*$ depends on the temperature
\cite{shag4,shag1} \begin{equation} M^*(T)\propto \frac{1}{\sqrt{T}}.
\end{equation}
The state of system with $M^*$ strongly depending on
$T$ and $r$ resembles the strongly correlated liquid.
In contrast to the strongly correlated liquid, there is no
energy scale $E_0$ given by Eq. (9). Such a system can be viewed
as a highly correlated liquid and  becomes the Landau Fermi
liquid at $T\to 0$. We expect that Eq. (9) is valid up to
temperatures $T\sim T_k\ll \varepsilon_F$  \cite{shag4,shag1}.

The LFL behavior is restored by the application of magnetic field
$B>B_{c0}$ and $T$-dependent corrections to the effective mass begin
with $T^2$-terms. Here $B_{c0}$ is a critical
field which suppresses the magnetically ordered state
and can be as big as 10-12 T \cite{bud} and
even bigger. If the
magnetically ordered state is absent then $B_{c0}=0$,
as it takes place in the case of CeRu$_2$Si$_2$ \cite{shag4,tak}.
At $r>0$ and $T^*(B)>T$, the
effective mass $M^*(B)$ of the restored LFL depends on magnetic field
$B$ \cite{shag4,shag1}
\beq M^*(B)\propto \frac{1}{(B-B_{c0})^{2/3}}.\eeq The
function $T^*(B)\propto (B-B_{c0})^{4/3}$ determines the line on the
$B-T$ phase diagram separating the region of the LFL behavior from the
NFL behavior taking place at $T>T^*(B)$ \cite{shag4,shag1}.  At $r<0$,
a recovery of the LFL behavior under applied magnetic fields takes
place at $T<T^*(B)\propto \sqrt{B-B_{c0}}$, while the effective mass is
given by \cite{pog} \beq M^*(B)\propto \frac{1}{\sqrt{B-B_{c0}}}.\eeq
It follows from Eqs. (11) and
(12) that the Kadowaki-Woods ratio $K=A(B)/\gamma^2(B)$ is
conserved because $A(B)\propto (M^*(B))^2$ and 
$\gamma(B)\propto M^*(B)$. Thus, the quasiparticle
systems described by Eqs.  (8), (9), (10), (11), and (12) determine the
universal behavior which is observed in measurements on HF metals
\cite{shag4}.

To capture and summarize the salient features of magnetic field-tuned
CQP observed recently in CeCoIn$_5$ \cite{bi,pag1}, we apply
the above consideration based on FCQPT. A recent study of CeCoIn$_5$
in magnetic fields $B>B_{c0}$ have
revealed that the coefficients $A(B)$ and $C(B)$,
describing scattering in the LFL regime and
determining the $T^2$ contributions to the resistivity $\rho$
and thermal resistivity $\kappa_r$ respectively, possess the same
critical field dependence
\beq A(B)\propto C(B)\propto\frac{1}{(B-B_{c0})^{4/3}}, \eeq
with $B_{c0}=5$T, so that the ratio $A(B)/C(B)=c$  \cite{pag1}. Here $c$
is a field-independent constant characterizing electron-electron
scattering in metals and having a typical value of $0.47$, see e.g.
\cite{ben,pag2}. The observed critical exponent $4/3$ is in excellent
agreement with that of given by Eq. (11) because $A(B)\propto
C(B)\propto (M^*(B))^2$. Such the parallel behavior of charge and heat
transport with the scattering rate growing as $T^2$
shows that the delocalized fermionic excitations are
the Landau quasiparticles carrying charge $e$.
We note that these should be destroyed in the case of CQPT
\cite{sac,voj}. Nonetheless, let us assume for a moment that
these survive. Since the heat and charge transport tend to
strongly differ in the presence of the critical fluctuations of
superconducting nature, the constancy of the ratio rules out the
critical fluctuations \cite{pag1}.  On the other hand, one could expect
that some kind of critical fluctuations could cause the observed
behavior. For example, large scattering from antiferromagnetic
fluctuations of finite momenta could degrade the heat and charge
transport in a similar way \cite{pag2}. In this case,
in order to preserve the Kadowaki-Woods ratio these
fluctuations are to properly influence the specific heat which
characterizes the thermodynamic properties of the system and is not
directly related to the transport one. On the other hand, there are no
theoretical grounds for this. Therefore, the conservation of the
Kadowaki-Woods ratio observed in recent measurements on CeCoIn$_5$
\cite{bi} definitely seems to rule out these fluctuations. While both
the constancy of Kadowaki-Woods ratio \cite{bi} and the constancy of
the $A(B)/C(B)$ ratio \cite{pag1} give strong evidence in favor of the
quasiparticle picture. We remark that the above consideration of
relationships between critical fluctuations and FCQPT is in agreement
with these facts.

It is instructive to briefly analyze the behavior of the system
when the magnetic field is changed through $B_{c0}$.
Broadly speaking, the magnetic field can be regarded as the tuning
parameter (like e.g. pressure or density) which produces CQP,
while $B_{c0}$ can be regarded as the metamagnetic field
at which FCQPT takes place. We remember that the behavior of
system at FCQPT is not determined by critical fluctuations,
therefore, the behavior is not disturbed by the proximity to
first order phase transitions which could take place near
the metamagnetic field.
At $B<B_{c0}$, we have to replace $(B-B_{c0})$ with $(B_{c0}-B)$ in
Eqs. (11) and (12) and in the formulas determining functions $T^*(B)$.
Then it follows from Eqs.  (11) and (12) that at $B=0$, the system in
question exhibits the LFL behavior at temperatures $T<T^*(B_{c0})$.  At
elevated magnetic field and when $B\to B_{c0}$ from below, the
temperature $T^*(B)$ is depressed, $T^*(B)\to 0$, showing the reverse
trend with respect to the case when $B>B_{c0}$, so that the area of the
LFL behavior vanishes. Therefore, it might be said that increasing
magnetic field $B$ suppresses the LFL behavior enhancing the NFL
behavior. As a result, in both cases $r<0$ and $r>0$, at the elevated
magnetic field with $B<B_{c0}$, the evolution of the effective mass
$M^*$, specific heat $C$, resistivity $\rho$, etc, show the opposite
trend with respect to the case when $B>B_{c0}$.  For example, at
increasing magnetic field $B\to B_{c0}$, the Sommerfeld coefficient
$\gamma$ becomes divergent at $T\to 0$.  We can conclude, that passing
through the critical field $B=B_{c0}$ leads to a sharp maximum in
$M^*(B)$, $\gamma(B)$, $\rho(B)$, $\chi(B)$, etc, see Eq. (11). While
the Kadowaki-Woods ratio is conserved when the system exhibits the LFL
behavior. Note that very near $B=B_{c0}$ the temperature $T^*(B)\to0$,
therefore at finite temperatures, the sharp maximum is substituted by a
relatively broad maximum which becomes sharper as $T\to0$.
A similar behavior to the described above was
observed in measurements on Sr$_3$Ru$_2$O$_7$ \cite{gri} and
CeIrIn$_5$ \cite{cap}. A detailed consideration of these items will be
published elsewhere.

A few remarks related to the high-$T_c$ superconductivity are
in order here. Let us discuss the situation in a finite
external field as it is done in the case of continues
phase transitions.
Switch on the pairing interaction $\lambda V({\bf p}_1,{\bf
p}_2)$ generating the pairing field
\beq\Delta({\bf p})=\lambda \int V({\bf p},{\bf p}_1)
\kappa({\bf p}_1)\frac{d{\bf p}_1}{(2\pi)^3}.\eeq
Here $\lambda$ is the coupling
constant. The pairing field can be considered as an external field
because the order parameter
$\kappa({\bf p})=\sqrt{(1-n_0({\bf p}))n_0({\bf p})}$
is determined by the strong Landau interaction rather then
by the weak pairing interaction. We consider a weak
coupling regime at which the occupation numbers of quasiparticles
$n_0({\bf p})$ are not disturbed by the pairing interaction
and given by Eq. (6).
The pairing field, being linearly coupled with the
order parameter, see Eq. (14), removes the system from its critical
point replacing FCQPT with the superconducting phase transition.
In response to it, the FC
plateau inclines with the slope being proportional to $\Delta_1$, while
the effective mass becomes finite $M^*_{FC}\propto 1/\Delta_1$
\cite{ms}.  In that case, it turns out that the maximum value of the
superconducting gap $\Delta_1$ is linear with respect to small
values of $\lambda$ \cite{ks,ksk,ks1} and can be as large as
$\Delta_1\sim 0.1\varepsilon_F$ and the transition temperature
behaves as $T_c\propto x(x_{FC}-x)$ where $x$ stands for the doping
level \cite{ms,ms1}. At finite temperatures $T\leq T_c$, the
quasiparticle excitations are the Bogoliubov quasiparticles.
Moving along this line, it is possible to explain the
main features of the high-$T_c$ superconductivity including a recovery
of the LFL behavior under applied magnetic fields as well, see e.g.
\cite{ms,ks,ksk,ms1,ms2,plam}.

As the Landau theory of Fermi liquid, the theory of the
high-temperature superconductivity based on FCQPT deals with the
quasiparticles which are elementary excitations of low energies.
This theory produces the general qualitative description of
the superconducting state, normal one and the recovery of the LFL
behavior under the application of magnetic field
\cite{ms,ks,ksk,plam}. On the other hand, one can choose the
phenomenological parameters and obtain the quantitative consideration
of the superconductivity as it can be done in the framework of the
Landau theory when describing a particular normal Fermi-liquid,
say liquid $^3$He. Thus,
any theory which is capable of describing FC and incorporates with the
BCS theory will produce the qualitative picture of the superconducting
state and the normal state which coincides with the picture based on
FCQPT.	Both of the pictures can agree at a numerical level provided
the corresponding parameters are adjusted.  For example, since the
formation of  flat band corresponding to FC
is possible in the Hubbard model \cite{irk},
one can, generally speaking,  repeat the results of the theory based on
FCQPT within the Hubbard model. It is appropriate mention here that the
corresponding numerical description confined to the case of $T=0$ has
been obtained within the Hubbard model \cite{rand,pwa}.

In conclusion, we have shown that
in contrast to CQPT, whose physics is dominated by thermal
and quantum fluctuations and characterized by the absence of
quasiparticles, the physics of a Fermi system near FCQPT or
undergone FCQPT is determined by quasiparticles
resembling the Landau quasiparticles.  Contrary to the Landau
quasiparticles, the effective mass of these quasiparticles is
strongly depends on the temperature, magnetic fields, density,
etc. This system of quasiparticles has the general properties
and determines the universal behavior of the Fermi system under
consideration including the recovery of the LFL behavior under applied
magnetic fields which preserves the Kadowaki-Woods ratio.
This universal behavior persists up to relatively
high temperatures comparatively to the case when  such a behavior is
determined by CQPT. We have analyzed the
striking recent measurements of specific heat, charge and heat
transport which where used to study the nature of magnetic
field-induced QCP in heavy-fermion metal
CeCoIn$_5$ and shown that the
observed facts are in good agreement with our scenario based on
FCQPT and certainly seem to rule out the critical fluctuations
related with CQPT. We have demonstrated that the Fermi system in
question can be represented by the electronic systems of the high-$T_c$
superconductors, HF metals and by some two-dimensional Fermi systems.
Finally, our general consideration suggests that FCQPT and the
emergence of novel quasiparticles at QCP and behind QCP and resembling
the Landau quasiparticles are qualities intrinsic to strongly
correlated substances.\\

The authors are grateful for the financial support provided
by the Korea Science and Engineering Foundation
through the Center for Advanced Plasma Surface Technology at
SungKyunKwan University.

\end{document}